\documentclass[12pt]{article}
\usepackage{graphicx}
\usepackage{epsf}
\usepackage{latexsym}
\newcommand{\be}{\begin{equation}}
\newcommand{\ee}{\end{equation}}
\newcommand{\ba}{\begin{array}{l}}
\newcommand{\ea}{\end{array}}
\newcommand{\bqa}{\begin{eqnarray}}
\newcommand{\eqa}{\end{eqnarray}}

\begin{document}

\begin{center}
{\Large\sf Further remarks on $\pi\pi$ scattering dispersion
relations}
\\[10mm]
{\sc Zhiguang Xiao and Hanqing Zheng\footnote{e-mail:
zhenghq@pku.edu.cn}}
\\[5mm]
{\it Department of Physics, Peking University, Beijing 100871,
P.~R.~China }
\\[5mm]
\begin{abstract}
The naive use of higher order perturbation theory leads the
left--hand cut integrals in $\pi\pi$ dispersion
relations~\cite{hjy,Xiao01} divergent. This problem is discussed
and solved.  Also we point out that the Adler zero condition
imposes three constraints on the dispersion relations. The $\sigma
$ pole position is determined using the improved method,
$M_\sigma= 483\pm 13 {\rm MeV}$,$\Gamma_\sigma=
   705 \pm 50{\rm MeV}$. The
scattering length parameter is found to be in excellent agreement
with the experimental result.
\end{abstract}
\end{center}
\\

In Ref.~\cite{hjy,Xiao01}, we have established a new dispersion
representation for the partial wave $\pi\pi$ scattering $S$
matrix. The key point for  setting up the dispersion
representation is the observation that both the real part and the
imaginary part of $S$ defined in the physical region are analytic
functions on the  cut plane expressed in terms of poles, the left
hand cut integrals and the kinematic factor,
$\rho=\sqrt{1-4m_\pi^2/s}$.
 To be specific, the real part of $S$, $\cos(2\delta_\pi)$ (where
$\delta_\pi$ is the scattering phase shift) and the imaginary part
of the $S$ matrix, $\sin(2\delta_\pi)$ satisfy the following
dispersion relations:
 \bqa
 \sin(2\delta_\pi)&\equiv&\rho F \ ,\nonumber \\ F(s)&=&
 \alpha+ \sum_i{\beta_i \over 2i\rho(s_i)(s-s_i)}-\sum_j{1\over
  2i\rho(z^{II}_j)S'(z^{II}_j)(s-z^{II}_j)}  \nonumber \\ &&
  +{1\over\pi}\int_{\rm L}{{\rm Im_L}F(s')
  \over s'-s} ds'+{1\over\pi}\int_{\rm R}{{\rm Im_R}F(s')
  \over s'-s} ds'\ ,
   \label{s2d}
  \eqa
  and
   \bqa
  && \cos(2\delta_\pi) \equiv\tilde{F}
   =\tilde{\alpha}+\sum_i {\beta_i\over2(s-s_i)}
  \nonumber\\
  &&+\sum_j{1\over
 2S'(z^{II}_j)(s-z^{II}_j)} +{1\over\pi}\int_{\rm L} {{\rm
 Im_L}\tilde F(s')
 \over s'-s} ds' +{1\over\pi}\int_{\rm R} {{\rm
 Im_R}\tilde F(s')
 \over s'-s} ds',  \label{c2d}
\eqa where  $s_i$ denote the possible bound state pole positions
and $\beta_i$ are the corresponding residues of $S$; $z_j^{II}$
denote the possible resonance pole positions on the second sheet.
The integrals denote the cut contributions, L=$(-\infty,0]$ is the
left hand cut ($l.$$h.$$c.$) and R starts from the $\bar KK$
threshold once the $4\pi$ cut are neglected, $\alpha$ and
$\tilde\alpha$ denote the subtraction constants, and one
subtraction to the integrals in above expressions is understood.
With these dispersion relations we can then generalize the single
channel unitarity relation $S^+S=1$,  which is only valid in the
single channel physical region when $s$ is real, to the whole
complex $s$ plane~\cite{hjy}:
$\cos^2(2\delta_\pi)+\sin^2(2\delta_\pi)=1$. The latter is
equivalent to the well known generalized unitarity condition in
quantum mechanics but was firstly discussed in field theory in
Ref.~\cite{hjy}.

The above method is valid for any  partial wave scattering. It is
however worth pointing out that in the scattering process with a
non-vanishing angular momentum, $J$, restrictions among parameters
should exist to ensure the threshold behavior,
$\delta_\pi(s)\propto k^{2J+1}$, where $k=\sqrt{s-4}$.
 In order to make use of Eqs.~(\ref{s2d}) and
(\ref{c2d}) in phenomenological discussions, a knowledge on the
$l.$$h.$$c.$ integrals is necessary. It is not very clear how to
calculate these $l.$$h.$$c.$ integrals in the    nonperturbative
scheme. Predictions on the left hand cuts from nonperturbative
models like the Pad\'e approximation are not always
trustworthy~\cite{ang}. Therefore results from chiral perturbation
theory (CHPT) are used in estimating these integrals via the
following formula,
 \bqa {\rm Im_L}F&=&2{\rm Im_L Re_R}T(s)=2{\rm Im_L}T- 2{\rm Re_L Im_R}T\ ,\label{IMLF}
\\{\rm Im_L} \tilde F&=&-2\rho(s){\rm Im_L Im_R}T(s)\ ,\label{tF}
\eqa since
 \be\label{FtF} F=2{\rm Re}_R T\ ,\,\,\,\tilde{F}=1-2\rho {\rm Im}_RT\ .\ee
In Eqs.~(\ref{IMLF}), (\ref{tF}) and (\ref{FtF}), $T$ is the
partial wave scattering $T$ matrix: $S\equiv 1+2i\rho T$.
 The quantity ${\rm Im_L}F$ is estimated at $O(p^4)$ in
 \cite{Xiao01} and ${\rm Im_L}\tilde F$ vanishes at this order.
 A
  question naturally arises as how close is the $O(p^4)$ results to
 the real situation. Since at $O(p^4)$ higher
 resonances do not
 contribute to ${\rm Im_L}F$ and ${\rm Im_L}\tilde F$,
 it is argued in \cite{hjy} that the contribution from
 the $t$ channel resonance exchange to the $l.$$h.$$c.$ is
 very small at low energies. It is confirmed by a calculation to
 the tree level $\rho$ exchange diagram~\cite{hjy2}, which
 indicates that the contribution to the $l.$$h.$$c.$ integral is
 numerically very small up to $s=1$GeV$^2$. Since resonance
 contributions saturate the $O(p^4)$ $\pi\pi$ interaction
 chiral Lagrangian \cite{rchpt} at low energies and they contribute ${\rm Im_L}F$ and ${\rm Im_L}\tilde F$ at
 $O(p^6)$, the above discussions  may suggest that the
 $O(p^4)$ results of ${\rm Im_L}F$ and ${\rm Im_L}\tilde F$ are
  good approximations to the real situations at $s< 1$GeV$^2$ despite of the ambiguity
  in choosing the cutoff parameter when estimating the integrals.
  However, even though high energy contributions to the left hand
  cut integrals may be small, the availability of the systematic use of
  perturbation theory in estimating ${\rm Im_L}F$ and ${\rm Im_L}\tilde
  F$ needs to be proved. This issue is not as trivial  as it looks like at first glance.

To have an understanding to the problem occurring in using higher
order perturbation theory results when estimating ${\rm Im_L}F$
and ${\rm Im_L}\tilde F$, let us focus on Eq.~(\ref{tF}). Since
${\rm Im_L}\tilde F$ vanishes at $O(p^4)$, the leading order
contribution is of $O(p^6)$,
 \bqa
 {\rm Im_L Im_R}T(s) &=&2\rho T_{2} {\rm Im_L}{\rm
 Re}_RT_{4}\nonumber\\
 &=&2\rho T_{2} {\rm Im_L}T_{4}-2\rho^2T_{2}^3\ .
 \eqa
 To obtain the second equation we have made use of the perturbative
unitarity relation, ${\rm Im}_RT_{4}=\rho T_{2}^2$ where (and
hereafter) the subscripts  denote the order of the chiral
expansions. Taking into account  the results from CHPT we find
that  when $s$ approaches $0_-$, ${\rm Im_L}\tilde F$ behaves as
$O((1/\sqrt{-s})^3)$ due to the presence of the kinematic factor.
The integration in the left hand integral in Eq.~({\ref{c2d}) is
therefore divergent when $s'\rightarrow 0_-$. If  higher order
results are used the problem is getting worse since there will be
higher powers of $1/\sqrt{-s}$. The same situation occurs in ${\rm
Im_L}F$ when using higher order results from ChPT.  We will
demonstrate in the following that this problem is only a deceptive
artifact inherited from perturbation theory and can be corrected.
In fact, from rather general considerations ${\rm Im_L}F(s)$ and
${\rm Im_L}\tilde{F}(s)$ should behave as $O(\sqrt{-s})$ and
$O(1/\sqrt{-s})$, respectively, when $s$ approaches zero. Hence
the left hand integrals in Eqs.~(\ref{s2d}) and (\ref{c2d}) are
well defined quantities.

In order to understand the behavior of ${\rm Im_L} \tilde{F}(s)$
when $s\rightarrow 0$, we should firstly understand the behavior
of $T^I_J(s)$ and ${\rm Im_L}T^I_J(s)$ as $s\rightarrow 0$. Since
$s=0$ is a branch point for $T^I_J(s)$ we first let $s$ approaches
0 from the positive side along the real axis. From the
partial wave projection formula %
\be T_J^I(s)={1\over
32\pi(s-4m_\pi^2)} \int^{0}_{4m_\pi^2-s} dt\, P_J(1+{2t\over
s-4m_\pi^2}) T^{I}(s,t,u)\ ,\,\,\, u=4m_\pi^2-s-t\ , \ee %
we conclude that the partial wave amplitude is regular at $s=0_+$,
since in the unphysical region $s=0$, $0\le t,u<4m_\pi^2$ the full
amplitude $T^I(s,t,u)$ contains no singularity. CHPT results are
consistent with the conclusion obtained from the general analysis:
actually in CHPT, at each order of the chiral expansion, the limit
$T^I_{J(2n)}(0_+)$ exist as demonstrated by the explicit
calculation for $n=1,2,3$ and it is reasonable to assume it exists
for arbitrary $n$. For the behavior of ${\rm Im_L}T^I_J(s)$ near
$s=0$, we have
 \be {\rm Im_L}{T^I_J(s)}={1+(-1)^{I+J}\over
32\pi(s-4m_\pi^2)} \int_{4m_\pi^2}^{4m_\pi^2-s} dt\,
P_J(1+{2t\over s-4m_\pi^2}){\rm Im}T^{I}_t(s,t)\,\,\, ; s\le 0\ ,
\ee
 from which we conclude ${\rm
Im_L} T^I_J(s)\sim(\sqrt{-s})^3$ as $s\rightarrow 0_-$: one factor
$s$ comes from the integral interval in the above equation,
another factor of $\sqrt{-s}$ comes from the threshold behavior of
${\rm Im}T^{I}_t(s,t)$ near $t=4m_\pi^2$. This observation is in
agreement with the $O(p^4)$ CHPT results in Ref.~\cite{Xiao01} and
can be obtained from a more careful discussion \cite{MartinBook}.
Again this behavior is expected to hold for ${\rm
Im_L}T^{I}_{J(2n)}$. For simplicity in the following discussion we
drop out the spin and isospin indices for partial wave amplitudes.

For the asymptotic behavior of ${\rm Im_R} T(s)$ as $s\rightarrow
0$, in perturbation theory the unitarity condition of the partial
wave amplitude is satisfied at each order of perturbation
expansion on the unitary cut $s>4m^2_\pi$: %
\bqa\label{pertUni} {\rm Im_R} T_{2}(s)&=&0\ , \nonumber \\
{\rm Im_R} T_{2n}(s)
&=&\rho(s)\sum_{i=1}^{n-1}T_{2n-2i}(s)T^*_{2i}(s)\, ,\ \mathrm{
for }\ n\geq 2\ .
\eqa %
To be specific, %
\bqa {\rm Im_R} T_4(s)&=&\rho(s)T_2^2(s)\, , \nonumber \\\nonumber\\%
{\rm Im_R} T_6(s)&=&2 \rho(s)T_2(s) {\rm Re_R}T_4(s)\nonumber \\%
&=&2 \rho(s)T_2(s)T_4(s)-2i\rho^2(s)T_2^3(s)  \, ,\nonumber\\\nonumber\\%
{\rm Im_R} T_8(s)&=& \rho(s)(2T_2(s){\rm Re_R}T_6(s)+T_4(s)T_4^*(s)) \nonumber \\%
&=& 2\rho(s)(T_2(s)T_6(s)+T_4^2(s))-6i\rho^2(s)T_2^2(s)T_4(s)-4\rho^3(s)T_2^4(s) \nonumber \\%
&\vdots&\nonumber\\
{\rm Im_R} T_{2n}(s)
&=&\rho(s)\sum_{i=1}^{n-1}T_{2n-2i}(s)T_{2i}(s)-
2i\rho(s)\sum_{i=1}^{n-1}T_{2n-2i}(s){\rm Im_R}T_{2i}(s). \eqa%
From above expressions we can easily figure out that as $s$
approaches 0  ${\rm Im_R} T_{2n}(s)$ is more and more singular
when we expand the amplitude to higher orders, because of the
higher power of  $\rho(s)$. We denote the most divergent term of
 ${\rm Im_R}T_{2n}(s)$ as ${\rm
Im_R}T_{2n}^{(1)}(s)$ and the next to leading divergent term as
${\rm Im_R}T_{2n}^{(2)}(s)$. By simple deduction, one  obtains for
$n\ge 3$ , \bqa {\rm
Im_R}T_{2n}^{(1)}(s)&=&(-2i\rho(s)T_2(s))^{n-2}\rho(s)T_2^2(s)
  \, ,\nonumber\\%
{\rm Im_R}T_{2n}^{(2)}(s)&=&(n-1)(-2i\rho(s)T_2(s))^{n-3}T_4(s)\rho(s)T_2(s) \, .
\eqa
 From the fact that $T_2(0)$ is only a nonzero real constant and ${\rm
Im_L}T_4(s)\sim O((\sqrt{-s})^3)$ as $s\rightarrow 0_-$, we
conclude that the behaviors of ${\rm Im_L Im_R} T_{2n}^{(1)}(s)$
and ${\rm Im_L Im_R} T_{2n}^{(2)}(s)$ as $s\rightarrow 0_-$ are,
\bqa {\rm Im_L Im_R}T_{2n}^{(1)}(s)\sim\left\{\matrix{
{\left(1\over \sqrt{-s}\right)}^{n-1}\,,{\rm\
 as\ } s\rightarrow0_-\ {\rm if\ n\ odd},\cr0\;\;\;\;\;\,,{\rm\
as\ } s\rightarrow0_-\ {\rm if\ n\ even},}\right.\nonumber\\
 {\rm Im_L
Im_R}T_{2n}^{(2)}(s)\sim\left\{\matrix{{\left(1\over
\sqrt{-s}\right)}^{n-5}\,, {\rm \ as\ } s\rightarrow0_-\ {\rm if\
n\ odd,}\cr {\left(1\over \sqrt{-s}\right)}^{n-2}\,, {\rm \ as\ }
s\rightarrow0_-\ {\rm if\ n\ even},}\right.\nonumber\\
 {\rm Re_L
Im_R}T_{2n}^{(1)}(s)\sim\left\{\matrix{0\;\;\;\;\;\,, {\rm  \ as\
} s\rightarrow0_-\ {\rm if\ n\ odd,}\cr {\left(1\over
\sqrt{-s}\right)}^{n-1}\,, {\rm \ as\ }
s\rightarrow0_-\ {\rm if\ n\ even},}\right.\nonumber\\
 {\rm Re_L
Im_R}T_{2n}^{(2)}(s)\sim\left\{\matrix{{\left(1\over
\sqrt{-s}\right)}^{n-2}\,, {\rm  \ as\ } s\rightarrow0_-\ {\rm if\
n\ odd,}\cr {\left(1\over \sqrt{-s}\right)}^{n-5}\,, {\rm \ as\ }
s\rightarrow0_-\ {\rm if\ n\ even}.}\right.\nonumber\\
\eqa To summarize, the most divergent term of ${\rm Im_L
Im_R}T_{2n}(s)$ and ${\rm Re_L Im_R}T_{2n}(s)$ are :
 \bqa
 {\rm Im_L
Im_R}T_{2n}(s)\sim\left\{\matrix{{\left(1\over
\sqrt{-s}\right)}^{n-1}\,, {\rm  \ as\ } s\rightarrow0_-\ {\rm if\
n\ odd,}\cr {\left(1\over \sqrt{-s}\right)}^{n-2}\,, {\rm \ as\ }
s\rightarrow0_-\ {\rm if\ n\ even}.}\right.\nonumber\\
 {\rm Re_L
Im_R}T_{2n}(s)\sim\left\{\matrix{{\left(1\over
\sqrt{-s}\right)}^{n-2}\,, {\rm  \ as\ } s\rightarrow0_-\ {\rm if\
n\ odd,}\cr {\left(1\over \sqrt{-s}\right)}^{n-1}\,, {\rm \ as\ }
s\rightarrow0_-\ {\rm if\ n\ even}.}\right. \eqa

From Eqs.~(\ref{IMLF}) and (\ref{tF}), ${\rm Im_L}\tilde{F}$ and
${\rm Im_L}F$ contain the same order of divergence as $\rho(s){\rm
Im_L Im_R}T(s)$ and ${\rm Re_L Im_R}T(s)$, respectively, as $s
\rightarrow 0$. For example, the most divergent term of ${\rm
Im_L}\tilde{F}$ at $O(p^6)$ behaves like $O((1/\sqrt{-s})^3)$ and
${\rm Im_L}F\sim O((1/\sqrt{-s}))$, when $s\rightarrow 0_-$ and
higher order divergences appear as higher order perturbation
expansions are used.

It is not difficult to demonstrate that the singular behavior of
${\rm Im_L}\tilde{F}$ and ${\rm Im_L}F$ at $s=0$ as discussed
above is just a spurious one inherited from the use of
perturbation theory.\footnote{Note that this problem will occur
not only in ChPT but also in all perturbation theories because
Eq.~(\ref{pertUni}) is correct in all perturbation theories and
all deductions above are based on this equation. In Pad\'e
approximations this problem  disappears.} To understand this we
notice that, for the complete, non-perturbative amplitude we have
 \be\label{unitary1} {\rm Im_R} T(s)= {\rho(s)T^2(s)\over
1+2i\rho(s)T(s)}\ , \ee
 which is obtainable from
the relation $T^*(s+i\epsilon)=T^{II}(s+i\epsilon)=T/S$ and
the single channel unitarity relation. As $s\rightarrow0$ we can
obtain from above equation that ${\rm Im_R} T(s)$ should not be
singular provided that $T(s)$ is not singular, or more precisely,
${\rm Re_R}T(0)=i{\rm Im_R}T(0)=T(0)/2$. Therefore, by simple
deduction using Eq.~(\ref{IMLF}) and ${\rm
Im_L}T\sim(\sqrt{-s})^3$, one can find that ${\rm
Im_L}\tilde{F}(s)$ is  $O((1/\sqrt{-s}))$ and ${\rm Im_L}F(s)$ is
$O(\sqrt{-s})$ as $s\rightarrow0_-$. The left hand cut integrals
are therefore well defined and are finite except at $s=0$ for
$\tilde{F}$.

The Eq.~(\ref{unitary1}) not only indicates the correct asymptotic
behavior of each quantity as $s\rightarrow 0$, but also the reason
why the naive use of perturbation results leads the left cut
integrals divergent -- if we could  sum up the perturbation series
to all orders the divergence problem would have disappeared:
\bqa {\rm Im_R} T&=&\sum_{n=2}^\infty {\rm Im_R} T_{2n} \nonumber \\%
&=&\sum_{n=2}^\infty\rho(s)\sum_{i=1}^{n-1}T_{2n-2i}(s)T^*_{2i}(s)\nonumber \\ %
&=&\rho(s)\sum_{i=1}^\infty\sum_{n=i+1}^\infty T_{2n-2i}(s)T^*_{2i}(s)\nonumber \\ %
&=&\rho(s)\sum_{i=1}^\infty\sum_{n=1}^\infty T_{2n}(s)T^*_{2i}(s)\nonumber \\ %
&=&\rho(s)\left(\sum_{i=1}^\infty T_{2i}(s)\right)^2-2i\rho(s)
\sum_{i=1}^\infty\sum_{n=1}^\infty T_{2n}(s){\rm Im_R}T_{2i}(s)\nonumber\\%
&=&\rho(s)\left(\sum_{i=1}^\infty T_{2i}(s)\right)^2-2i\rho(s)
\left(\sum_{n=1}^\infty T_{2n}(s)\right){\rm Im_R} T\ ,
\eqa %
from which we can deduce %
\bqa\label{SumImRT} {\rm Im_R}T&=&\sum_{n=2}^\infty {\rm Im_R}
T_{2n}=\frac{\rho(s)\left(\sum_{i=1}^\infty
T_{2i}(s)\right)^2}{1+2i\rho(s)\left(\sum_{i=1}^\infty
T_{2i}(s)\right)} \, .\eqa %
 Notice that in deriving Eq.~(\ref{SumImRT}) we do not need to
 use the knowledge on analytic continuation.
  Comparing with Eq.~(\ref{unitary1}), the above result is obtained by the
simple substitution of  $T$ expanded to all orders in perturbation
theory into Eq.~(\ref{unitary1}). As we discussed  earlier, the
chiral expansion of $T(0)$ is assumed to be well-defined since
each $T_{2n}(0)$ is finite and $\sim O((m^2_\pi/4\pi f^2_\pi)^n)$.
Therefore Eq.~(\ref{SumImRT}) suggests the correct way of
extracting ${\rm Im_L}\tilde{F}$ and ${\rm Im_L}F$ near $s=0$,
from a finite order perturbative calculation. That is to use
\be\frac{\rho(s)\left(\sum_{i=1}^n
T_{2i}(s)\right)^2}{1+2i\rho(s)\left(\sum_{i=1}^n
T_{2i}(s)\right)}\label{SumImRT'}\ee
 instead of $\sum_{i=2}^n {\rm Im_R} T_{2i}$.
The former expression removes the spurious divergence in the
latter introduced by the kinematic factor at $s=0$. The
Eq.~(\ref{SumImRT'}) works well in the vicinity of $s=0$, in other
places it should be considered equally well or equally bad as the
naive expression of perturbation expansion. In our early studies
we made the naive use of $O(p^4)$ CHPT to estimate ${\rm
Im_L}F(s')$ which gives an incorrect $O({1\over \sqrt{-s'}})$
behavior near $s'=0$ even though the left hand integral in
Eq.~(\ref{s2d}) is still definable except at $s=0$. However, the
numerical influence to the cut integral at $s\ge 4m_\pi^2$ is very
small since the integral interval overwhelmed by the spurious
divergence is very small. Hence  numerical results in estimating
the left hand integrals are only affected very little by the naive
use of perturbation theory.

In Ref.~\cite{hjy} we have discussed the constraint of the Adler
zero condition on the $\pi\pi$ scattering dispersion relation.
There we approximately fix the Adler zero at $s=m_\pi^2/2$ in the
I=J=0 channel and force the partial wave $S$ matrix being equal to
1 at $s=m_\pi^2/2$. However the method is not good enough since,
first of all, the Adler zero position for the partial wave
amplitude is not exactly  located at $s=m_\pi^2/2$, and secondly,
the  Adler zero condition actually impose more constraints than
what is previously considered.

In partial wave amplitudes, the position of Adler zero can not be
exactly given  because the Adler zero is defined at
$s=u=t={m_\pi^2}$ in the full amplitude and after the partial wave
projection one cannot fix its exact position.
Taking I=J=0 channel for example,  one can find a zero  at
$s=m_\pi^2/2$ in the tree level amplitude $T_2={2s-m_\pi^2\over
32\pi f_\pi^2}$. If the perturbative amplitude is a  good
approximation ($i.e.$, converges rapidly) to the real situation in
the vicinity of $s=m_\pi^2/2$ then the Adler zero position for the
partial wave amplitude, $s_A$, may exist and may be determined by
solving the equation $0=T(s_A)\simeq T_2(s_A)+T_4(s_A)+ ...$ using
the iteration method: $s_A\simeq m_\pi^2/2-16\pi
f_\pi^2(T_4(m_\pi^2/2)+...)$. For the given perturbative
amplitudes from CHPT the Adler zero position can be estimated
numerically: $s_A=0.419\pm 0.058$ in unit of $m_\pi^2$. The error
bar appeared in the estimate  reflects the uncertainties of
coupling constants of the chiral Lagrangian~\cite{Bijnens}. So in
the following we will not fix the position of Adler zero and
instead we use it as a parameter in our fit procedure.

In Ref.~\cite{hjy} we discussed the role of the Adler zero
condition in the global fit. What we did is to enforce $S$ in
Eq.~(9) of Ref.~\cite{hjy} being unity at the zero. Which,
however, did not make the full use of the Adler zero condition.
According to Eqs.~(\ref{FtF}), (\ref{unitary1}) and ${\rm
Re_R}T(s)=T(1+S)/(2S)$ it is easy to realize that
 \be
\label{AdlerCon}F(s_A)=0, \; \tilde{F}(s_A)=1, \;
\tilde{F}'(s_A)=0\ , \ee
 which therefore impose three constraints on the $\pi\pi$
 scattering dispersion relations. The Eq.~(\ref{AdlerCon}) is
 helpful in the phenomenological discussions. For example, it can
 be used to make one more subtraction to the dispersion integrals
 in Eq.~(\ref{s2d}) and two more subtractions to the dispersion
 integrals in Eq.~(\ref{c2d}). This is appreciable since more
 subtractions can in principle reduce the uncertainties when estimating
 those integrals, which mainly come from high energies.
 The  Eqs.~(\ref{s2d}) and (\ref{c2d}) can for example be recasted as:%
\bqa \label{F} F(s) &=&(s-s_A){2a^0_0\over
4m_\pi^2-s_A}+(s-4m_\pi^2)(s-s_A)\sum_j {i
/(2\rho(z^{II}_j)S'(z^{II}_j))\over
(z^{II}_j-4m_\pi^2)(z^{II}_j-s_A)(s-z^{II}_j)}
\nonumber\\&&+{(s-4m_\pi^2)(s-s_A)\over\pi}\int_{L+R} ds'{{\rm Im}
F(s')\over (s'-4m_\pi^2)(s'-s_A)(s'-s)}\ , \nonumber\\
\tilde{F}(s) &=&1+(s-s_A)^2(s-4m_\pi^2)\sum_j
{1\over2S'(z^{II}_j)(z^{II}_j-s_A)^2(z^{II}_j-4m_\pi^2)(s-z^{II}_j)}
\nonumber\\&&+{(s-4m_\pi^2)(s-s_A)^2\over\pi}\int_{L+R} ds'{{\rm
Im} \tilde{F}(s')\over (s'-4m_\pi^2)(s'-s_A)^2(s'-s)}\ ,\eqa
 where $a_0^0$ denotes the scattering length parameter. Here
 $a_0^0$ is no longer a free parameter since it is determined by
 $F(s_A)=0$ where $F$ is the one originally defined in Eq.~(1).

Using the improved dispersion relations as described above we
repeat the fit made in Ref.~\cite{hjy}. The fit procedure is the
same as before (for example, here we also take  $\epsilon=0.02$
which constrains the violation of unitarity ) except that $s_A$ is
no longer held fixed. Taking into account all the uncertainties
and variations of parameters we arrive at the following
results,\footnote{in a previous version, the fit was performed by
inappropriately taking $a_0^0$ being free, and the results were
given in Phys. Lett. {\bf B549}(2002)362 (Erratum). The fit
results are nevertheless not very sensitive to the different
treatment of the $a_0^0$ parameter.}
   \bqa\label{adler2}
  && M_\sigma= 483\pm 13 {\rm MeV}\ ,\,\,\, \Gamma_\sigma=
   705 \pm 50{\rm MeV}\ ;\nonumber\\
&&a_0^0= 0.223\pm 0.006\ ,\,\,\, s_A\simeq (0.268 - 0.309
)m_\pi^2.
  \eqa

Note that the scattering length parameter is now in excellent
agreement with the result of Ref.~\cite{newke4,CGL01}. The above
results should replace those given in Ref.~\cite{hjy}. Notice that
the error bars given above only represent the uncertainties from
our theoretical input and does not have a statistical meaning. The
position of the Adler zero is about $0.3m_\pi^2$ according to our
fit, which is not very far from the result from chiral
perturbation theory. The results on $f_0(980)$ are also similar to
those previously obtained. It is worth noticing that the new
results are very close to the results given in Table~2 of
Ref.~\cite{Xiao01} where the scattering length is constrained by
hand using the result of Ref.~\cite{newke4}. It was found that the
$\sigma$ pole position is rather sensitive to the scattering
length parameter~\cite{Xiao01}, the correct use of the Adler zero
condition is therefore  crucial in obtaining the correct
scattering length parameter and the $\sigma$ pole position.

We would like to thank Guang-You Qin for helpful discussions. This
work is supported in part by China National Nature Science
Foundation under grant number 10047003 and 10055003.

\end{document}